\documentclass{jltp}

\usepackage{graphicx}
\usepackage{amsmath}
\usepackage{amssymb}

\title{Zero--point vacancies in quantum solids}
\author{M. Rossi, E. Vitali, D. E. Galli and L. Reatto}
\address{Dipartimento di Fisica, Universit\`a degli Studi di Milano,\\
           via Celoria 16, 20133 Milano, Italy}

\runninghead{M. Rossi, E. Vitali, D.E. Galli and L. Reatto}
            {Zero--point vacancies in quantum solids}

\begin{document}

\maketitle

\begin{abstract}
A Jastrow wave function (JWF) and a shadow wave function (SWF) 
describe a quantum solid with Bose--Einstein condensate; i.e. a supersolid.
It is known that both JWF and SWF describe a quantum solid with also a 
finite equilibrium concentration of vacancies $x_v$.
We outline a route for estimating $x_v$ by 
exploiting the existing formal equivalence between the absolute square of the 
ground state wave function and the Boltzmann weight of a classical solid. 
We compute $x_v$ for the quantum solids described by JWF and SWF employing very
accurate numerical techniques.
For JWF we find a very small value for the zero point vacancy concentration,
$x_v=(1.4\pm0.1)\times10^{-6}$.
For SWF, which presently gives the best variational description of solid $^4$He,
we find the significantly larger value $x_v=(1.4\pm0.1)\times10^{-3}$ 
at a density close to melting.
We also study two and three vacancies.
We find that there is a strong short range attraction but the vacancies do not
form a bound state. 
\end{abstract}

\section{INTRODUCTION}

A fascinating topic in modern quantum many--body Physics is the possibility of a 
supersolid state of matter, i.e. a solid which displays some superfluid properties.
Since the first theoretical speculations on its existence in the early seventies,
supersolid has gained a wide attention.\cite{meisel} 
Being such a supersolid phase a direct consequence of macroscopic manifestation of 
quantum properties, the highly quantum solid $^4$He was soon recognized to be the 
natural candidate to realize it; but it has revealed to be elusive to the 
experimental observation until 2004, when the observation of non classical rotational
inertia effects (NCRI)\cite{chan} provided the first possible signature of its 
presence.
Many experiments followed the original one,\cite{chan,indep} giving rise to a
puzzling scenario: NCRI is now observed in torsional oscillator experiments both
when solid $^4$He is inside a porous material as well as in bulk solid $^4$He, but 
the detected supersolid fraction $\rho_s$ shows a strong dependence on the details of
the experimental realization.
The observed dependence of $\rho_s$ on the sample, on $^3$He concentration\cite{3he} 
and also on annealing\cite{chan,indep,reppy2} leads to a largely diffuse conviction 
that NCRI should be essentially ascribed to extrinsic disorder rather than to an 
intrinsic property of solid $^4$He.
Some aspects of the measurements (as the metastability at low temperatures, the 
value of the critical velocity, $\rho_s$ which vanishes in a continuous way with a 
dissipation peak, the dependence of $\rho_s$ on the oscillator frequency) seem to 
point in the direction of a liquid of vortices as suggested by Anderson.\cite{ander2}
On the other hand, recent experiments with an high quality $^4$He crystal\cite{clark}
and the observation of a peak in the specific heat in correspondence of the 
supersolid transition\cite{clark2} seem to put some restrains to the interpretation 
of NCRI as purely extrinsic effect.
There is no doubt that defects play a relevant role in NCRI, but what happens if a 
more and more perfect crystal is grown?
Will the supersolid response disappear or some NCRI effect survives? 
The answers to these questions lie in the properties of the ground state wave 
function of a quantum crystal.
To what extent particles localization induced by the crystallization process 
succeeds in overcoming the zero--point motion of the atoms?  
Is it possible that even in the true equilibrium state a finite amount of intrinsic 
defects, such as vacancies,\cite{anderson,galn} may survive leading to a superfluid
response?
This question concerning the true nature of the ground state of solid $^4$He is 
still open.\cite{galn,bonin2}
Zero--point vacancies, i.e. vacancies in the ground state of solid $^4$He, were the 
first proposed mechanism for the supersolid state.\cite{andr,ches}
Delocalized defects, acting as a dilute Bose gas, undergo a Bose--Einstein 
condensation (BEC) resulting in superfluid--like properties of the crystal.
But zero--point vacancies have never been experimentally observed, however actual 
experiments cannot exclude them at a concentration lower than about 
0.4\%.\cite{simm2}
Our understanding of the solid phase of $^4$He is far from being complete and a 
full microscopic description that accounts for the whole supersolid phenomenology is
still lacking.

Even if it is possible to obtain essentially exact results for most properties of
solid $^4$He by means of quantum simulations, e.g. PIMC\cite{ceperley} at finite 
temperature and SPIGS\cite{spigs} at zero temperature, they do not offer a simple 
straight way to answer the question of the true nature of the ground state.
Such computations give compelling evidence that the commensurate crystal, i.e. a 
crystal in which the number of particles is equal to the number of lattice sites, 
has no BEC.
Much more delicate is the question of the presence of defects like vacancies in the
ground state.
Whether solid $^4$He, or any other possible quantum crystal, should have a
commensurate or incommensurate ground state is a fundamental question due to the
important role of such defects on the occurrence of BEC.
Quantum simulation of solid $^4$He with one vacancy\cite{pede,gal1,bonin2} indicates
that the energy of this state is above the one of a perfect crystal and this has
been taken as evidence that no ground state vacancies are present in solid $^4$He.
However it has been argued that such quantum simulations present in literature do
not allow to directly answer the question.\cite{galn}
In fact, in Monte Carlo simulations of crystals, the small number of particles, the
use of periodic boundary conditions and commensuration effect of the simulation box
with the lattice structure impose a constraint that makes impossible, in practice,
for the system to develop an equilibrium concentration of vacancies.\cite{andersen}
This holds not only for canonical and micro-canonical computations (where it should
be somehow obvious since the number of particles and the volume of the system are
fixed quantities during the simulation), but also for grand canonical
simulations.\cite{andersen}
Furthermore the equilibrium concentration could be so small to easily have escaped
detection in simulations of the ground state of solid $^4$He.\cite{anderson}
Therefore the equilibrium concentration of zero point vacancies $x_v$ can be
obtained only indirectly, by a statistical thermodynamical analysis of an extended
system, exactly as for classical solids.\cite{frenkel}

On the other hand we know two model system that have zero point defects and BEC.
This is the case of the quantum solid described either by a Jastrow wave function 
(JWF) or by a shadow wave function (SWF).
Both these wave functions have a more or less successful long career as trial wave 
functions in variational descriptions of solid as well as liquid $^4$He.
It is well known that at large densities such wave functions describe a crystalline solid.
Moreover, the presence of BEC in such solid phase is stated by a theorem both for 
JWF\cite{reatto} and for SWF,\cite{mass,gal1} and in the latter case the condensate 
fraction has recently been estimated.\cite{gal2}

Furthermore, by following the Chester's argument for supersolidity,\cite{ches} it is 
possible to infer that such translationally invariant wave functions (JWF and SWF)
give rise to a finite concentration of vacancies.
Even though this argument is common knowledge, no estimate exists of how large is 
such equilibrium concentration of vacancies, except for an approximate calculation 
done some years ago for JWF.\cite{still}
In this paper we have developed the route outlined by Chester and we have exploited
the most advanced numerical techniques to estimate the equilibrium concentration of 
vacancies $x_v$ in the ground state of the quantum solids described by JWF and SWF.
By a full thermodynamical analysis of the equivalent classical crystal, we have 
estimated $x_v=(1.4\pm0.1)\times10^{-6}$ for a JWF and $x_v=(1.4\pm0.1)\times10^{-3}$
for the SWF at a density close to the melting one.

The paper is organized as follows: in Section~\ref{sec:meth} we introduce the 
Jastrow and the shadow wave functions briefly reporting also the Chester's argument 
for vacancy induced supersolidity; a route for the estimate of the equilibrium
concentration of vacancies $x_v$ is outlined in Section~\ref{sec:eqxv}, which deals
also with the numerical tools involved in the computation.
Our result for JWF and SWF $x_v$ are reported in Section~\ref{sec:res} and discussed
in Section~\ref{sec:con} which contains also our conclusions.

\section{VARIATIONAL WAVE FUNCTIONS}
\label{sec:meth}

The ground state wave function for a system of $N$ bosons must be symmetric under
permutations of the coordinates, and it can be chosen as a real non negative 
function.
The simplest choice for the wave function in the interacting bosons case is 
represented by the Jastrow wave function
\begin{equation}
 \label{jwf}
 \psi_J(R)=\frac{1}{\sqrt{Q_N}}\prod_{i<j}e^{-\frac{1}{2}u(r_{ij})}\quad,
\end{equation}
where $u(r)$ accounts for the direct correlations among two particles and $Q_N$ is 
the normalization constant.
This wave function is translationally invariant and describes a liquid or a solid 
depending on the density.
The solid phase comes as a spontaneously broken symmetry effect at large density.
Moreover such a wave function has been proved to have a finite Bose--Einstein 
condensate fraction even in the solid phase.\cite{reatto}
There is a formal identity, recognized long ago,\cite{mcmi} between $|\psi_J|^2$ 
and the normalized probability distribution in configurational space of $N$ 
classical particles at an inverse temperature $\beta$ interacting via a potential 
\begin{equation}
 \label{Uj}
 \mathcal{U}_{\rm J}(R)=\frac{1}{\beta}\sum_{i<j}u(r_{ij})\quad.
\end{equation}
In addition, if as commonly assumed $u(r)$ vanishes at large distance, the 
normalization constant $Q_N=\int dR\prod_{i<j}e^{-u(r_{ij})}$ is equal to the
canonical configurational partition function of this classical system, so that the 
logarithm of $Q_N$ turns out to be proportional to the excess Helmholtz free energy.
The Chester's argument in favor of the vacancy induced supersolidity lies on this 
formal analogy.\cite{ches}
At large density this equivalent classical system corresponding to $|\psi_J|^2$ is 
a crystalline solid.
For this equivalent classical system the equilibrium concentration of vacancies 
$x_v$($=(M-N)/M$ where $M$ is the number of lattice sites) is non--zero even if a 
single vacancy has a finite cost of local free energy and this raises from the gain 
in configurational entropy.\cite{kittel}
It is then straightforward to infer that also the quantum solid described by 
\eqref{jwf} should have the same $x_v$.

One comment is in order.
The pseudopotential $u(r)$ should be determined by minimization of the expectation
value of the Hamiltonian.
In the liquid phase the quantitative results given by the JWF are reasonable even 
if not very accurate.\cite{mcmi}
Much less so it is in the solid phase.
In fact, with JWF one gets solidification at a density much larger than the 
experimental value and the particles turn out to be very localized.\cite{hans}
A wave function that greatly improves over the JWF is the so called shadow wave 
function.
In fact, SWF presently gives the best variational representation of the ground state
properties of $^4$He both in the liquid and in the solid phase.\cite{moro}

In a SWF the atoms are correlated not only by standard direct correlations among 
particles, but also indirectly via the coupling to a set of subsidiary (shadow) 
variables.\cite{viti}
Integration over the shadows variables introduces effective interparticle 
correlations between pair of atoms but also between triplets and, in principle, to 
all higher orders in an implicit way; this is done so efficiently that it is 
possible to treat the liquid and the solid phase with the same functional 
form\cite{moro,viti} without the need to introduce {\it a priori} equilibrium 
positions to localize the atoms around lattice positions. 
Even for SWF in fact, solidification emerges as a result of a spontaneously broken 
translational symmetry due to the increased correlations as the density of the 
system increases.
Moreover local disorder processes due to zero point motion are in principle allowed.
This makes the use of such a wave function specially useful in the present context,
where our purpose is to investigate the ground state nature of a quantum crystal.

The general form of a SWF is given by
\begin{equation}
 \label{swf}
 \Psi(R)=\frac{1}{\sqrt{Q_N}}\phi_r(R)\int\!dS\ \Theta(R,S)\phi_s(S),
\end{equation}
where $R=({\bf r}_1,\dots,{\bf r}_N)$ and $S=({\bf s}_1,\dots,{\bf s}_N)$ are,
respectively, the real and the shadow coordinates of the $N$ atoms and $Q_N$ is the
normalization constant.
As usual with SWF,\cite{viti} $\phi_r(R)$ is a Jastrow function which gives the 
direct correlations between particles, and we assume a McMillan form for the 
pseudopotential:\cite{mcmi}
\begin{equation}
 \label{phir}
 \phi_r(R) = \prod_{i<j}e^{-\frac{1}{2}\left(\frac{b}{r_{ij}}\right)^m}\qquad.
\end{equation}
$\phi_s(S)$ is another Jastrow function giving the inter shadow correlations; as 
pseudopotential we use the rescaled and shifted interatomic potential\cite{viti}:
\begin{equation}
 \label{phis}
 \phi_s(S) = \prod_{i<j}e^{-\delta V(\gamma s_{ij})}\qquad.
\end{equation}
$\Theta(R,S)$ is a Gaussian kernel coupling each shadow variable to the 
corresponding real one:
\begin{equation}
 \label{kernel}
 \Theta(R,S)= \prod_{i=1}^Ne^{-C|{\bf r}_i-{\bf s}_i|^2}\qquad.
\end{equation}

When SWF is employed as trial wave function in a variational description of $^4$He,
the parameters $b$, $m$, $\delta$, $\gamma$ and $C$ have to be determined by 
minimizing the expectation value of the Hamiltonian
\begin{equation}
 \label{ham}
 \hat H = -\frac{\hbar^2}{2m}\sum_{i=1}^N\nabla^2_i + \sum_{i<j}V(r_{ij})
\end{equation}
where $V$ is the helium--helium interatomic potential.
For instance when an hcp commensurate crystal with $N=180$ $^4$He atoms at 
$\rho=0.0293$\AA$^{-3}$, slightly above the melting density, is considered and the 
standard Aziz potential\cite{aziz} is chosen as $V$, the minimization of \eqref{ham}
gives the values $b=2.76$\AA, $m=5$, $\delta=0.11$K$^{-1}$, $\gamma=0.875$ and 
$C=0.8725$\AA$^{-2}$ as optimal values for the variational parameters.\cite{viti}
Also SWF belong to the class of wave functions that display BEC even in the solid 
phase,\cite{mass} and a recent investigation computed the condensate fraction to be
$5\times10^{-6}$ at the melting density.\cite{gal2}

When performing ground state calculations with a SWF by the Monte Carlo methods, due
to the integration over shadow variables, two independent kinds of shadow particles 
must be taken into account in performing averages, as it can be easily seen, for 
example, by writing the expectation value of a diagonal observable $O(R)$
\begin{equation}
 \langle O \rangle = \int dRdSdS' p(R,S,S') O(R)
\end{equation}
where the probability density $p(R,S,S')$ is given by
\begin{equation}
 \label{prob}
 p(R,S,S') = \frac{1}{Q_N}\phi_s(S')\Theta(R,S')\phi_r(R)^2\Theta(R,S)\phi_s(S)
\end{equation}
\begin{equation}
 \label{Qswf}
 Q_N = \int dRdSdS' \phi_s(S')\Theta(R,S')\phi_r(R)^2\Theta(R,S)\phi_s(S) \quad.
\end{equation}

A quantum--classical correspondence exists for SWF as well.\cite{mass}
In fact, there is a formal analogy between Eq.~\eqref{prob} and the probability 
distribution density in configurational space of $N$ classical triatomic molecules 
with flexible bonds at an inverse temperature $\beta$ interacting via the potential
\begin{equation}
 \label{swfecs}
 \begin{split}
  \mathcal{U}_{\rm S}(R,S,S')& = 
   \frac{1}{\beta}\left[\sum_{i<j}\left(\frac{b}{r_{ij}}\right)^m
    +C\sum_{i=1}^N\left(|{\bf r}_i-{\bf s}_i|^2+|{\bf r}_i-{\bf s'}_i|^2\right)\right.\\
   & \left.+\delta\sum_{i<j}\left(V(\gamma s_{ij})+V(\gamma s'_{ij})\right)\right]\quad.
 \end{split}
\end{equation}
Each molecule is then composed by one real (central) particle and two shadows.
In this quantum--classical analogy no interaction other than the intramolecular 
forces corresponding to the real--shadow coupling and the intermolecular forces 
between particle of the same species (provided by the real--real, shadow--shadow and
shadow$'$--shadow$'$ pseudopotentials) are present. 
The statistical sampling of $|\Psi(R)|^2$ then, maps the quantum system of $N$ 
particles in an equivalent classical system of $N$ flexible triatomic molecules with
special interactions and the logarithm of the normalization constant \eqref{Qswf} is
proportional to the excess free energy of such classical molecular system.
Furthermore the Chester's argument can be extended also to this molecular solid to 
prove that also SWF gives rise to a finite equilibrium concentration of vacancies.

\section{EQUILIBRIUM CONCENTRATION OF VACANCIES}
\label{sec:eqxv}

As already pointed out, Chester's argument\cite{ches} ensures that in the quantum 
solids described by JWF and SWF (i.e. the quantum solids whose ground state wave 
functions are JWF and SWF) have a finite equilibrium concentration of vacancies, 
whose amount obviously depends on the chosen pseudopotential parameters.
The problem of the estimate such equilibrium concentration of vacancies $x_v$ in 
the quantum solids at $T=0$ K is then formally equivalent to the computation of 
$x_v$ in their equivalent classical solids.
The presence of a finite concentration of vacancies at equilibrium conditions means 
that the partition function $Q_\mathcal{N}$ of a macroscopic system of $\mathcal{N}$
particles has contributions from different pockets in configurational space 
corresponding not only to the commensurate state but also to states with a different 
number of vacancies; and the pockets with vacancies give the main 
contribution.\cite{galn}
This translates in the quantum case to say that the wave function $\psi(R)$ of a 
bulk system is describing at the same time both states with and without vacancies, 
and that the overwhelming contribution to the normalization constant $Q_\mathcal{N}$ 
of the macroscopic system derives from pockets corresponding to a finite concentration
of vacancies.\cite{galn}

Only about 10 years ago, $x_v$ has been computed for solid $^4$He by exploiting this 
quantum--classical isomorphism\cite{still} and choosing as wave function a JWF 
with a McMillan pseudopotential\cite{mcmi}.
At melting density $x_v$ turned out to be $x_v=6.36\times10^{-6}$.
This is interesting but one should keep in mind that JWF gives a very poor description 
of solid $^4$He,\cite{hans,moro} moreover the pseudopotential parameters were chosen to 
stabilize the solid and not to minimize the expectation value of the quantum 
Hamiltonian at the chosen density.
In addition, a quasi--harmonic approximation was used to estimate the free energy 
cost of a vacancy but the accuracy of this approximation is not known.
Here we improve that estimate by employing direct simulations method to compute the
involved free energies without any harmonic approximation.
More important, with these methods we compute $x_v$ also for the SWF, a wave function 
that gives a very accurate description of solid $^4$He.\cite{moro}

\subsection{Classical solid}

Pronk and Frenkel\cite{frenkel} outlined a full grand canonical route to estimate the
equilibrium concentration of vacancies in the thermodynamic limit for a classical solid.
The grand partition function $\mathcal{Z}$ of the crystal accounts for the fluctuations
both of the number of particles $N$ and of the number of lattice sites $M$.
The canonical partition function of the crystal with $n=M-N$ vacancies and $M$ lattice
sites is factorized, under the assumption of non--interacting defects, as
\begin{equation}
 \label{binomiale}
 Q_{M-n}(V,T)\simeq\frac{M!}{n!(M-n)!}Q^{(n)}(V,T)
\end{equation}
where $Q^{(n)}(V,T)$ is now the partition function of the crystal with $n$ fixed 
vacancies at given lattice positions and the binomial factor accounts for all the 
possible configurations of such vacancies.
This assumption allows to recognize in $\mathcal{Z}$ a binomial expansion over $n$ that
can be replaced with its sum value.
The sum over the lattice site number $M$ is approximated with a standard maximum term 
argument and then the obtained concentration of vacancies at the thermodynamic 
equilibrium is given by\cite{frenkel} 
\begin{equation}
 \label{xv}
  x_v= e^{-\beta(\mu-f_1)}
\end{equation}
where $\mu$ is the chemical potential of the defect--free crystal and $-f_1$ is the
change in the free energy of the crystal due to the creation of a vacancy at a 
specific lattice position.

At a first sight, the formula \eqref{xv} for $x_v$ seems to be obtained under the 
assumption of localized defect, and then should not be used in the case of mobile
vacancies.
This is not true, since the canonical partition function $Q_{M-n}$ can be always 
factorized over the configurations of the defects even if mobile.\cite{howard}
In fact, the partition function $Q_{M-n}$ is made of static integrations over those 
particle configurations which are compatible with the presence of $n$ vacancies.
The corresponding volume in the configurational space can be split in sub-volumes
each corresponding to a particular configuration of the $n$ defects, i.e. $Q^{(n)}$, 
independently on how much time the defects spend in any configuration.
When vacancies are not interacting all these $Q^{(n)}$ are equal and 
Eq.\eqref{binomiale} follows.

Exploiting thermodynamic relations, the exponent in \eqref{xv} can be written as
$-\beta(\mu-f_1) = (M-1)\beta(f_0-f_d)-\beta P/\rho$; then, in order to compute 
$x_v$, we need the free energy per particle of the perfect, $f_0$, and of the 
defected (i.e. with one vacancy), $f_d$, crystal and of the pressure $P$.
The pressure $P$ for a classical solid is quite straightforwardly obtained with the 
virial method\cite{fbook,allen} or from volume perturbations.\cite{volum}
Convergence of both the methods to the same value would provide a check on the 
estimated pressure.
The computation of the free energy (both $f_0$ and $f_d$) is less immediate: free 
energy cannot be directly obtained by a single Monte Carlo simulation.\cite{fbook}
We recall that the free energy of the equivalent classical solid is related to the 
logarithm of the normalization constant $Q_N$ for the quantum crystal which, also,
is never explicitly computed.

Let us consider a classical system with interparticle potential energy $U(R)$ where 
$R$ denotes the set of coordinates.
The thermodynamic integration method gives a way for reconstructing free energy 
differences by integration over a reversible path in the phase space.\cite{fbook}
In a simulation we are not limited to use a thermodynamic path, rather all the 
parameters in the potential energy function can be used as thermodynamic variables.
Then, in order to obtain the absolute free energy $F$, we must have at one end of 
the path a system whose free energy is known (reference system), and to the other 
end we have the system of interest.
For solids a popular method is the one of Frenkel and Ladd\cite{ladd,polson}. 
The basic idea is to reversibly transform the solid under consideration into an 
Einstein crystal.
To this end, the atoms are harmonically coupled to their lattice sites.
Let $\mathcal{U}_{\rm Ein}(R)$ be the Einstein crystal chosen as reference system
and let us consider the potential 
\begin{equation}
 \label{ulambda}
 \mathcal{U}_\lambda(R) = (1-\lambda)\mathcal{U}(R) +\lambda \mathcal{U}_{\rm Ein}(R)\quad.
\end{equation}
As $\lambda$ varies from 0 to 1 the system described by $\mathcal{U}_\lambda$ goes 
from the system of interest ($\lambda=0$) to the reference one ($\lambda=1$).
It can be easily found that\cite{fbook}
\begin{equation}
 \left.\frac{\partial F(\lambda)}{\partial\lambda}\right|_{N,V,T} = 
  \left\langle\frac{\partial\mathcal{U}_\lambda(R)}{\partial\lambda}\right\rangle_\lambda
\end{equation}
and then, for the free energy
\begin{equation}
 \label{freeen}
 F = F_{\rm Ein} + \int_0^1d\lambda\ \left\langle\frac{\partial\mathcal{U}_\lambda(R)}{\partial\lambda}\right\rangle_\lambda =
     F_{\rm Ein} + \int_0^1d\lambda\ \left\langle\mathcal{U}_{\rm Ein}(R) - \mathcal{U}(R)\right\rangle_\lambda   
\end{equation}
where $\left\langle\dots\right\rangle_\lambda$ denotes the canonical average for a 
system with potential energy function $\mathcal{U}_\lambda(R)$.
It should be noted that the thermodynamic integration method \eqref{freeen}
is intrinsically static, i.e. the derivative of the free energy is obtained in a 
sequence of equilibrium simulations each for different values of the coupling 
parameter $\lambda$.
The potential energy for the reference Einstein crystal is given by 
\begin{equation}
 \label{Uein}
 \mathcal{U}_{\rm Ein}(R) = \sum_{i=1}^N\alpha_i\left({\bf r}_i - {\bf r}_i^0\right)^2
\end{equation}
where ${\bf r}_i^0$ are the lattice positions; its free energy is
\begin{equation}
 F_{\rm Ein} = -\frac{3}{2\beta}\sum_{i=1}^N\log\frac{\pi}{\alpha_i\beta}
\end{equation}
where, $\beta$ is the inverse temperature.
In the case of the defected crystal, a particle is removed  keeping the simulation 
box volume and the lattice parameter unchanged.
The spring constants $\alpha_i$ can be adjusted to optimize the accuracy of the 
numerical integration in \eqref{freeen} and are usually chosen to make the 
mean--squared displacement equal for $\lambda=0$ and $\lambda=1$.\cite{fbook}

\subsection{Quantum solid}

In the quantum system $U(R)$ corresponds to the so called pseudopotentials, for 
instance $U(R)$ of the equivalent classical solid corresponds to $\log|\psi_J(R)|^2$
in the case of a JWF.
The method outlined in the previous subsection for computing $x_v$ might be used also 
for a quantum crystal described by a JWF or a SWF. 
However some modifications have to be introduced; due to the peculiar characteristics 
of the considered classical solids whose thermal equilibrium mimics the zero--point 
motion of quantum crystals, particles (or molecules) can be mobile.
This is particularly true when a vacancy is present.
In such situation the harmonic coupling to lattice positions would produce unbounded
fluctuations because of the sudden suppression of exchange processes, which are 
allowed by the wave function, when $\lambda\ne0$.
A trick to bypass this problem is to break the thermodynamic path in two steps.
In the first the equivalent classical crystal is changed into another crystal in 
which, by acting on the variational parameters, the pseudopotential are made so 
tight that the zero point motion is strongly reduced and exchange processes are
greatly suppressed.
In the second step this new crystal is transformed into the reference Einstein crystal 
constructed as prescribed by the Frenkel--Ladd method.\cite{fbook}
We have already underlined that in thermodynamic integration all the parameters in
the potential function can be chosen as parameter to define the thermodynamic 
path.\cite{fbook} 

By construction the Frenkel--Ladd method gives access to $Q^{(n)}$; in fact, the vacancy 
localization is explicitly introduced via the coupling to the Einstein crystal where it
has a fixed position.
This is true also for our two step thermodynamic path except than for the very beginning 
of the first step, where the vacancy is able to move.
Strictly speaking this would introduce a systematic error, but in practice this error is
smaller than the statistical one when only one vacancy is present.
This has been verified by considering statistically independent estimates of $\beta f_d$
which turn out to be compatible within the statistical error.
In fact, thanks to the periodic boundary conditions the most of the sampled configurations
(i.e. the ones where the vacancy occupies a lattice site) can be rigidly translated to keep
the vacancy in the same original position without affecting the computed averages.

We have checked our numerical code for the estimate of the free energy in classical 
systems with the Frenkel--Ladd method by reproducing the results in 
Ref.~\onlinecite{polson} for a system of soft spheres, both with a single direct 
thermodynamic path from the interest crystal to the Einstein crystal, and with a two 
steps thermodynamic path as described above.
It is known that the Frenkel--Ladd method is sensitive to the size scaling.\cite{polson}
We have considered then different boxes of increasing size (different number of lattice 
sites $M$) and we have found, for all the considered crystals the correct linear 
dependence of $\beta f + \log M/M$ on $1/M$.\cite{polson} 
Since both $\beta f_0$ and $\beta f_d$ displays the same scaling law, the value of
$\beta(\mu-f_1)$, and then of $x_v$, shows no appreciable dependence on $M$.

\section{RESULTS}
\label{sec:res}

\subsection{Equilibrium concentration of zero--point vacancies for Jastrow wave function}

\begin{figure}[t]
 \begin{center}
  \includegraphics[width=10cm]{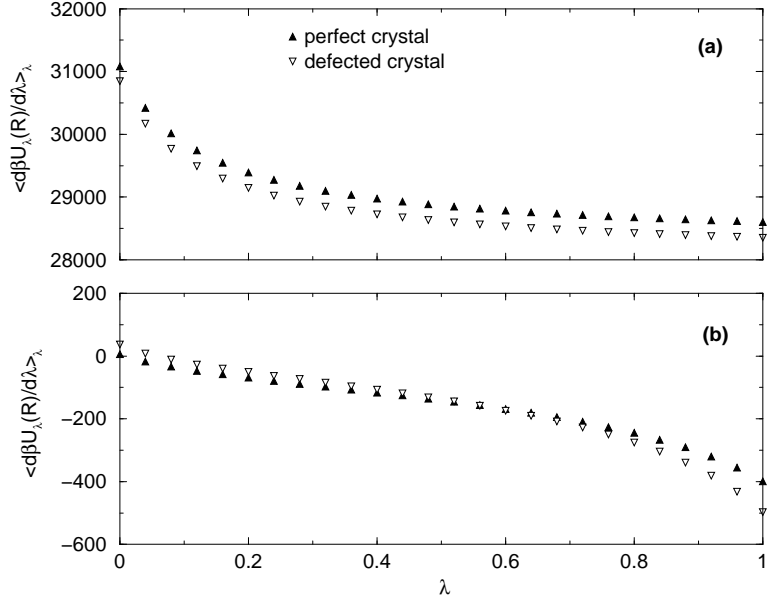}
 \end{center}
 \caption{\label{f:jwf} Obtained values for 
          $\langle\partial\beta\mathcal{U}_\lambda(R)/\partial\lambda\rangle_\lambda$
          in a JWF fcc crystal with $M=256$ at $\rho = 0.029$\AA$^{-3}$.
          Error bars are smaller than the used symbols.
          (a) first step: $\mathcal{U}_\lambda(R)=b_\lambda\mathcal{U}_{\rm J}(R)$ 
          with $\mathcal{U}_{\rm J}(R)$ as defined in Eq.~\eqref{Uj}, 
          $b_\lambda=(1-\lambda)b+\lambda(b_f/b)^m$ and $b_f=6.0$\AA.
          (b) second step: $\mathcal{U}_\lambda(R)=(1-\lambda)\mathcal{U}_{\rm J'}(R)
          +\lambda\mathcal{U}_{\rm Ein}(R)$ where the harmonic constants of the Einstein
          crystal are chosen such that $3/2\beta\alpha=0.0318$\AA$^2$ in the perfect 
          crystal and 0.0355\AA$^2$ in the defected one.}
\end{figure}

In a first computation we have considered the same wave function studied in 
Ref.~\onlinecite{still}, i.e. we have considered a JWF \eqref{jwf} with a 
McMillan\cite{mcmi} pseudopotential to describe a fcc crystal in a box with $M=256$ at
$\rho = 0.029$\AA$^{-3}$.
The pseudopotential parameters are $b=4.238$\AA~and $m=6$.
As already noticed this $b$ value does not minimize the expectation value of the 
Hamiltonian, but it is chosen so that the solid phase is stable at the considered 
density.
To avoid any vacancy mobility problem when computing $f_d$, the thermodynamic path
has been split into two steps.
The first step runs from the JWF crystal (i.e. the classical crystal equivalent to 
$|\psi_J(R)|^2$) to a JWF$'$ crystal obtained by increasing by a factor 1.4 the 
parameter $b$; this reduce the Lindemann ratio from $r_L=0.15$ to $r_L=0.05$, which
is small enough to ensure vacancy localization.
The second step links this JWF$'$ crystal to an Einstein crystal of the type 
described by Eq.~\eqref{Uein} with the same spring constant for all the particles.
In order to ensure a good convergence of the results, we have considered at least 26
values of the coupling parameter for each step and for each value of the coupling
parameter as many as $2.5\times10^5$ sweeps are produced at the equilibrium.
The obtained results for $\langle\partial\beta\mathcal{U}_\lambda(R)/\partial\lambda\rangle_\lambda$
are reported in Fig.~\ref{f:jwf} both for the first and the second thermodynamic
path steps.
The final result is $-\beta(\mu-f_1) = -13.476 \pm 0.057 $ which corresponds to an 
equilibrium concentration of vacancies $x_v=(1.4\pm0.1)\times10^{-6}$.
Our result is about 5 times smaller than the previous estimate 
$x_v=6.36\times10^{-6}$ of Ref.~\onlinecite{still}.
This discrepancy is presumably due to the fact that in the older estimate, the free 
energy cost of a vacancy was computed relying on an harmonic approximation.
Notice that the value of $x_v$ has an exponential dependence on $-\beta(\mu-f_1)$, 
and it is easy to see that a value of $-\beta(\mu-f_1)$ 11.2\% higher than the one
obtained in our computation would give the same $x_v$ value as Ref.~\onlinecite{still}.

$x_v$ for JWF turns out to be very small.
This is not too surprising because we know that a JWF can describe a solid but the 
particles turn out to be much more localized than in $^4$He.

\subsection{Equilibrium concentration of zero--point vacancies for shadow wave function}

The outlined method for the evaluation of the logarithm of the normalization constant 
$Q_N$ can be extended also to the molecular classical solid corresponding to the SWF.
A convenient way to build the Einstein crystal for such SWF crystal is to couple
each particle of a ``molecule'' (in the present case two shadows and one real) to the 
lattice position with an harmonic spring.\cite{anwar}
Since the choice of the pseudopotential parameters in a variational theory is 
related to a minimization principle, we choose as parameters the optimal ones to
describe solid $^4$He at $\rho = 0.0293$\AA$^{-3}$, a density which is slightly above
melting in order to reasonably prevent instabilities.

\begin{figure}[t]
 \begin{center}
  \includegraphics[width=10cm]{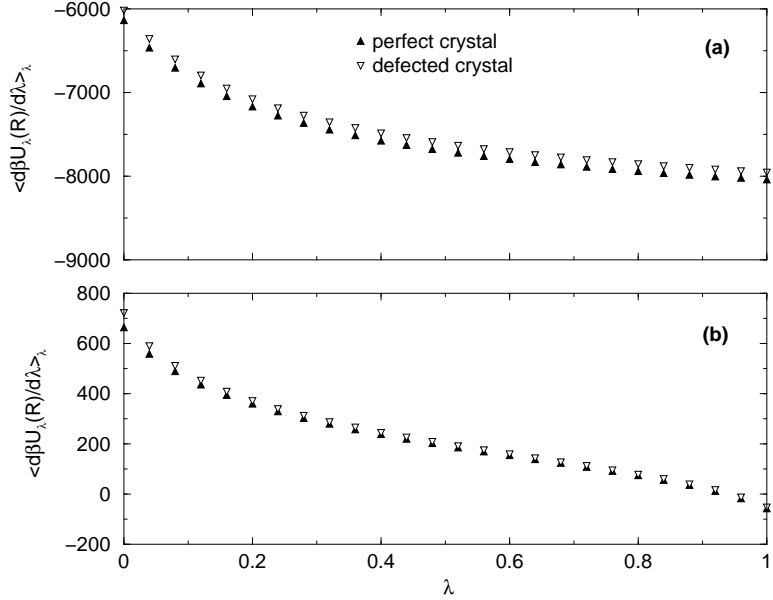}
 \end{center}
 \caption{\label{f:swf} Obtained values for 
          $\langle\partial\beta\mathcal{U}_\lambda(R)/\partial\lambda\rangle_\lambda$
          in a SWF hcp crystal with $M=180$ at $\rho = 0.0293$\AA$^{-3}$.
          Error bars are smaller than the used symbols.
          (a) first step: $\beta\mathcal{U}_\lambda(R)=\sum_{i<j}\left(b/r_{ij}\right)^m
          +C_\lambda\sum_{i=1}^N\left(|{\bf r}_i-{\bf s}_i|^2+|{\bf r}_i-{\bf s'}_i|^2\right)
          +\delta_\lambda\sum_{i<j}\left(V(\gamma s_{ij})+V(\gamma s'_{ij})\right)$
          with $C_\lambda=(1-\lambda)C+\lambda C_f$, $C_f=5.0$\AA$^2$, 
          $\delta_\lambda=(1-\lambda)\delta+\lambda\delta_f$ and $\delta_f=0.5$K$^{-1}$.
          (b) second step: $\mathcal{U}_\lambda(R)=(1-\lambda)\mathcal{U}_{\rm S'}(R)+
          \lambda\mathcal{U}_{\rm Ein}(R)$ where the harmonic constants of the 
          Einstein crystal are chosen such that $3/2\beta\alpha=0.0766$\AA$^2$ 
          both in the perfect and in the defected crystal.}
\end{figure}

We have considered a simulation box which houses a hcp crystal with $M=180$ lattice
positions and we have considered a two steps thermodynamic path.
The first step starts from an optimized SWF (i.e. a crystal wave function where the 
variational parameters are set to minimize the expectation value of the quantum 
Hamiltonian\cite{viti}) and ends on a SWF$'$ crystal, obtained by increasing the 
real--shadow correlations (parameter $C$) by a factor 6, and the shadow--shadow 
correlations (parameter $\delta$) by a factor 4.5 in \eqref{swfecs}.
In this way the Lindemann ratio of the SWF crystal reduces to $r_L=0.11$ from the
starting $r_L=0.27$.
The second step links this new SWF$'$ crystal to the reference one: the Einstein 
crystal.
Even in this case we have considered at least 26 values of the coupling parameter 
for each thermodynamic integration step.
Our results for $\langle\partial\beta\mathcal{U}_\lambda(R)/\partial\lambda\rangle_\lambda$
are reported in Fig.~\ref{f:swf} for both the thermodynamic path steps.
We have found $-\beta(\mu-f_1) = -6.56 \pm 0.07 $; and then the estimated 
equilibrium concentration of vacancies for a SWF at the melting density is
$x_v=(1.4\pm0.1)\times10^{-3}$.

We have checked our result by following different thermodynamic paths.
We have considered different SWF$'$ crystals, provided that the variational 
parameter were chosen in such a way to reduce the Lindemann ratio to values lower or
comparable to those of standard classical solids.
We have taken into account also two different kind of reference Einstein crystals: 
one with the same spring constant for all the particles and one with different 
spring constants for shadow and real particles.
The same $x_v$ value is recovered in all the considered cases within error bars.

The obtained $x_v$ for SWF is not so far from the actual experimental bound to the 
zero--point vacancies concentration in $^4$He of 0.4\%.\cite{simm2} 
Then improvements in experimental investigations could answer the question if SWF 
describes well also the equilibrium concentration of vacancies in the ground state.
Moreover, by using the BEC transition temperature $T_{\rm BEC}$ for an ideal Bose
gas with mass equal to the vacancy effective mass as obtained with SWF\cite{gal1}
($m_v=0.35m_{He}$) as an estimate of the supersolid transition temperature, we
obtain $T_{SS}\simeq 11.3 x_v^{2/3} = 141$ mK, which is about 2.3 times larger than
the experimental estimated transition temperature $T = 60$ mK.\cite{clark}
One possible origin of this discrepancy is the fact that $x_v$ is estimated via a 
variational technique.
As we shall discuss later, the outlined route to compute $x_v$ can be in principle 
extended to exact techniques.
Another origin is the presence of a significant vacancy--vacancy interaction.

\subsection{Vacancy--vacancy correlations}

Eq.~\eqref{xv} for $x_v$ is valid under the assumption of non interacting defects.
A key aspect to be investigated is whether in the quantum solid described
by SWF vacancies interact and then if a dilute gas of vacancies is stable.
Therefore we have studied systems with two and three vacancies.
Here we report the results for a system of $N=537$ atoms (plus twice as much 
shadows) at density $\rho = 0.0293$\AA$^{-3}$ in a box with periodic boundary 
condition such that an hcp crystal with $M=540$ lattice positions fits in.
Starting from an initial configuration corresponding to  such hcp lattice in which
3 particles have been removed, we find that the crystalline state is stable and
the periodicity is such that the number of density maxima is equal to 540, i.e.
3 more of the 537 atoms.
This means that we have an hcp crystal with 3 mobile vacancies.
We have monitored the 3 vacancies positions during the Monte Carlo sampling, and by 
histogramming their relative distances on at least $4\times10^6$ equilibrium 
configuration we have constructed a vacancy--vacancy correlation function 
$g_{\rm vv}(r)$.
\begin{figure}[t]
 \begin{center}
  \includegraphics[width=9cm]{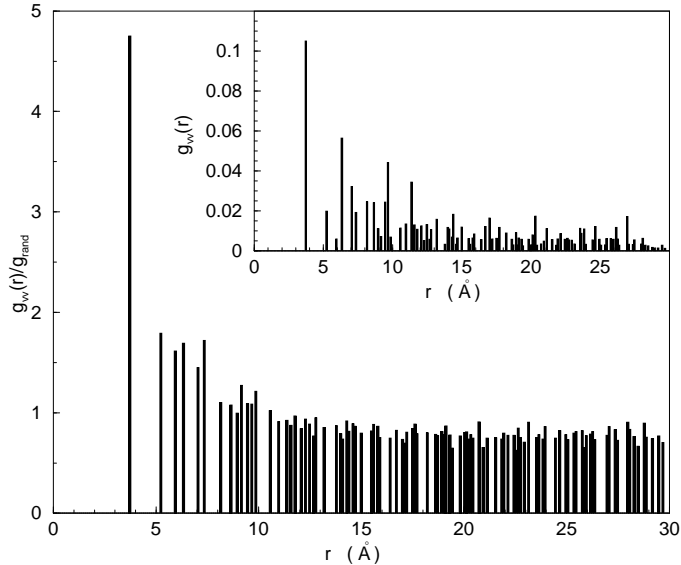}
 \end{center}
 \caption{\label{f:corr} Vacancy-vacancy correlation function $g_{\rm vv}(r)$
          computed in an hcp crystal at the melting density containing three
          vacancies and $M=540$ lattice positions.
          $g_{\rm vv}(r)/g_{\rm rand}$ is the vacancy-vacancy correlation function
          normalized with the correlation function of an ideal gas of 3 particles
          on the same hcp lattice, while in the inset $g_{\rm vv}(r)$ is reported
          normalized to unity.}
\end{figure}
In the inset of Fig.~\ref{f:corr}, we plot the obtained $g_{\rm vv}(r)$ normalized to 
unity, which gives the probability, given one vacancy at the origin,
to find another one at distance $r$.
In the main of Fig.~\ref{f:corr} we report $g_{\rm vv}(r)$ normalized with the 
correlation function of an ideal gas of 3 particles on the same hcp lattice.
In this case we have a vacancy pair distribution function.
The determination of the vector positions of the vacancies in a crystalline 
configurations is far from being trivial due to the large zero point motion, to the 
vacancy being mobile and because the center of mass is not fixed. 
Our algorithm proceeded as follows: given a configuration explored by the Metropolis
sampling, we find the crystal lattice which best fits the positions of the real 
particles; this is achieved through the construction of a ``Gaussian local density'' 
model  which is optimized adjusting the position of the center of mass of the lattice 
so that the ``molecules'' (a real and two shadow variables) fall in high density 
regions. 
Then, the position of a vacancy is defined as that site which does not host any 
atom in its Wigner-Seiz cell and whose nearest neighbors are not double 
occupied.\cite{pede}

From the plot of $g_{\rm vv}(r)$ (Fig.\ref{f:corr}), we see that the vacancies
explore the whole available distance range. 
We may so conclude that no sign of bound state is detected, at least
one the scale of the simulation box,
and it seems quite improbable a larger range bound state.
Similar results is found for two vacancies.
The observed large value for $g_{\rm vv}(r)$ at nearest neighbor distance is an 
indication of some short range attraction. 
But this interaction seems not large enough to give a bound state at least for up to 
3 vacancies, as evident in the tail of $g_{\rm vv}(r)/g_{\rm rand}$ which approaches 
the unity, i.e. vacancies behave almost like ideal particles for distances beyond about 
10 \AA.
We then conclude that the vacancy gas should be stable in bulk SWF solid; and, for 
very low concentrations, as the expected equilibrium one, the approximation of 
vacancies as non interacting defects seems to be reasonable.
However, the effect of an attractive interaction between vacancies, if not strong 
enough to lead to phase separation, would affect the equilibrium 
concentration of vacancies.
For example, the concentration of divacancies is given by
\begin{equation}
 \label{xv2}
 \frac{x_{2v}}{x_v^2}=\frac{q}{2}e^{-\beta\Delta f}
\end{equation}
where $q$ is the coordination number of the considered lattice ($q=12$ for hcp and 
fcc) and $\Delta f$ is the gain in the free energy in bringing two single vacancies 
together to form a pair.\cite{howard}
The relation \eqref{xv2} is the one expected from the application of the mass action 
law to the vacancy association reaction.\cite{howard}
The overall effect is that the total concentration of vacancies should somewhat
increase with respect to the value computed in the previous Section. 

\section{CONCLUSIONS}
\label{sec:con}

We have presented the first quantitative estimate of the concentration of vacancies
for a quantum solid described by the SWF which has been variationally optimized to
describe solid $^4$He.
We have computed $x_v$ exploiting the formal analogy between the ground state of 
bosons and the Boltzmann weight for classical particles and using advanced simulation
methods.
At melting we find $x_v=(1.4\pm0.1)\times10^{-3}$.
This is a sizable concentration not too far from the present experiment upper bound for
zero point vacancies in solid $^4$He.
$x_v$ for SWF is much larger than the value $x_v=(1.4\pm0.1)\times10^{-6}$ given by a
JWF. 
This, presumably, is due to the much stronger localization of particles described by a
JWF compared to SWF.
The study of two and three vacancies in the hcp crystal within the SWF shows the 
presence of a significant short range attraction but such multiple vacancies do not form
a bound state.

SWF has been proved to be so successful in describing many properties of solid $^4$He
that it actually represents the best available variational description of $^4$He.
As all the variational descriptions it suffers of some limitations, for example SWF
gives a finite condensate fraction also for a commensurate crystal,\cite{gal2} which 
turns out to be zero when  computed with exact techniques such as PIMC at finite 
temperature\cite{prok,cepe} and SPIGS at zero temperature.\cite{note} 
It has been suggested that this defect of SWF might be due to a qualitative
lack in the functional form of the wave function, which, probably, misses the 
description of zero point motion of excitations other than longitudinal phonons, 
such as the transverse ones.\cite{note}
For this reason some prudence is needed with respect to the relevance of the obtained 
concentration of ground state vacancies for SWF to the real ground state of solid 
$^4$He.
For example, in Ref.~\onlinecite{galn} it was argued that if the true ground state has 
a finite concentration of vacancies, the energy per particle computed in a box with 
the correct $x_v$ should be lower than the one computed for the commensurate crystal.
When the Aziz potential\cite{aziz} is used as interaction potential for solid $^4$He
in the quantum Hamiltonian \eqref{ham}, this property is unfulfilled by SWF at their
equilibrium concentration of vacancies.
This will be true only if the energy per particle is computed with the potential 
that has the SWF as exact ground state in \eqref{ham}.
Unfortunately such a potential is unknown; while it is known the one that has the JWF
as exact ground state.\cite{still} 
It is easy to check, as pointed out on a general discussion in Ref.~\onlinecite{cepe2}, 
that this is not a simple pair potential but it has a very strong 3-body term, which is 
about 1.3 times the 2-body one in the considered case.

The outlined route for the estimate of the equilibrium concentration of vacancies 
can, in principle, be extended also to the exact SPIGS\cite{spigs} method which 
recovers more and more accurate approximations of the true ground state wave function
by successive short time projections in imaginary time starting from a SWF.
The classical solid equivalent to SPIGS turns out to be a collection of open polymers
whose length (number of constituent monomers) is fixed by the number of considered
projection steps.
The Chester's argument applies also to this polymeric crystal which is expected to have
an equilibrium concentration of vacancies.
Further investigation with exact technique is not trivial but advisable.

This work was supported by the INFM Parallel Computing Initiative, by Supercomputing
facilities of CILEA and by Mathematics Department ``F. Enriques'' of the Universit\`a
degli Studi di Milano.


\begin{thebibliography}{99}

\bibitem{meisel} M.W. Meisel, Physica B {\bf 178}, 121 (1992);
\bibitem{chan} E. Kim and M.H.W. Chan, Nature {\bf 427}, 225 (2004);
               J. Low Temp. Phys. {\bf 138}, 859 (2005);
               Science {\bf 305}, 1941 (2004);
               Phys. Rev. Lett. {\bf 97}, 115302 (2006).
\bibitem{indep} A.S.C. Rittner and J.D. Reppy, Phys. Rev. Lett. {\bf 97}, 165301 (2006);
                M. Kondo, S. Takada, Y. Shibayama and K. Shirahama, J. Low Temp. Phys. {\bf 148}, 695 (2007);
                A. Penzev, Y. Yasuta and M. Kubota, J. Low Temp. Phys. {\bf 148}, 677 (2007);
                Y. Aoki, J.C. Graves and H. Kojima, Phys. Rev. Lett. {\bf 99}, 015301 (2007).
\bibitem{3he} E. Kim, J.S. Xia, J.T. West, X. Lin, A.C. Clark and M.H.W. Chan, Phys. Rev. Lett. {\bf 100}, 065301 (2008).
\bibitem{reppy2} A.S.C. Rittner and J.D. Reppy, Phys. Rev. Lett. {\bf 98}, 175302 (2007):
                 J. Low Temp. Phys. {\bf 148}, 671 (2007).
\bibitem{ander2} P.W. Anderson, Nature Physics {\bf 3}, 160  (2007).
\bibitem{clark} A.C. Clark, J.T. West and M.H.W. Cahn, Phys. Rev. Lett. {\bf 99}, 135302 (2007).
\bibitem{clark2} X. Lin, A.C. Clark and M.H.W. Cahn, Nature {\bf 449}, 1025 (2007).
\bibitem{anderson} P.W. Anderson, W.F. Brinkman and D.H. Huse, Science {\bf 310}, 1164 (2005).
\bibitem{galn} D.E. Galli and L. Reatto, Phys. Rev. Lett. {\bf 96}, 165301 (2006).
\bibitem{bonin2} M. Boninsegni, A.B. Kuklov, L. Pollet, N.V. Prokof'ev, B.V. Svistunov and M. Troyer,
                 Phys. Rev. Lett. {\bf 97}, 080401 (2006).
\bibitem{andr} A. F. Andreev and I. M. Lifshitz, Soviet Phys. JETP {\bf 29}, 1107 (1969).
\bibitem{ches} G. V. Chester, Phys. Rev. A {\bf 2}, 256 (1970).
\bibitem{simm2} R.O. Simmons and R. Blasdell, APS March Meeting 2007, Denver, Colorado.
\bibitem{ceperley} D.M. Ceperley, Rev. Mod. Phys. {\bf 67}, 279 (1995).
\bibitem{spigs} D.E. Galli and L. Reatto, Mol. Phys. {\bf 101}, 1697 (2003);
                J. Low Temp. Phys. {\bf 134}, 121 (2004).
\bibitem{pede} F. Pederiva, G.V. Chester, S. Fantoni and L. Reatto, Phys. Rev. B {\bf 56}, 5909 (1997).
\bibitem{gal1} D.E. Galli and L. Reatto, J. Low Temp. Phys. {\bf 124}, 197 (2001).
\bibitem{andersen} W.C. Swope and H.C. Andersen, Phys. Rev. A {\bf 46}, 4539 (1992).
\bibitem{frenkel} S. Pronk and D. Frenkel, J. Phys. Chem. B {\bf 105}, 6722 (2001).
\bibitem{reatto} L. Reatto, Phys. Rev. {\bf 183}, 334 (1969).
\bibitem{mass} L. Reatto and G.L. Masserini, Phys. Rev. B {\bf 38}, 4516 (1988).
\bibitem{gal2} D.E. Galli, M. Rossi and L. Reatto, Phys. Rev. B {\bf 71}, 140506(R) (2005).
\bibitem{cepe2} B.K. Clark and D.M. Ceperley, Phys. Rev. Lett. {\bf 96}, 105302 (2006).
\bibitem{still} J. A. Hodgdon and F. H. Stillinger, J. Stat. Phys. {\bf 79}, 117 (1995).
\bibitem{mcmi} W. L. McMillan, Phys. Rev. {\bf 138}, A442 (1965).
\bibitem{kittel} C. Kittel, {\it Introduction to solid state physics} (Wiley, New York, 1976).
\bibitem{hans} J.P. Hansen and E.L. Pollock, Phys. Rev. A {\bf 5}, 2651 (1972).
\bibitem{moro} S. Moroni, D.E. Galli, S. Fantoni and L. Reatto, {\it Phys. Rev. B} {\bf 58}, 909 (1998).
\bibitem{viti} S. A. Vitiello, K.Runge and M.H. Kalos, Phys. Rev. Lett. {\bf 60}, 1970 (1988);
               T. MacFarland, S.A. Vitiello, L. Reatto, G.V. Chester and M.H. Kalos,
               Phys. Rev. B {\bf 50}, 13577 (1994).
\bibitem{aziz} R.A. Aziz, V.P.S. Nain, J.S. Carley, W.L. Taylor and G.T. McConville, J. Chem. Phys. {\bf 70}, 4320 (1979).
\bibitem{howard} R.E. Howard and A.D. Lidiard, Rep. Prog. Phys. {\bf 27}, 161 (1964).
\bibitem{fbook} D. Frenkel and B. Smith, {\it Understanding Molecular Simulations}, 2nd Ed.
                (Academic Press, London, 2002).
\bibitem{allen} M.P. Allen and D.J. Tildesley, {\it Computer Simulations of Liquids}
                (Oxford University Press, Oxford, 1990).
\bibitem{volum} E. de Miguel and G. Jackson, J. Chem. Phys. {\bf 125}, 164109 (2006).
\bibitem{ladd} D. Frenkel and A.J.C Ladd, J. Chem. Phys. {\bf 81}, 3188 (1984).
\bibitem{polson} J.M. Polson, E. Trizac, S. Pronk and D. Frenkel, J. Chem. Phys. {\bf 112},
                 5339 (2000).
\bibitem{anwar} J. Anward, D. Frenkel and M.G. Noro, J. Chem. Phys. {\bf 118}, 728 (2003).
\bibitem{maham} G.D. Maham and H. Shin, Phys. Rev. B {\bf 74}, 214502 (2006).
\bibitem{prok} M. Boninsegni, N. Prokof'ev and B. Svistunov, Phys. Rev. Lett. {\bf 96}, 105301 (2006).
\bibitem{cepe} B.K. Clark and D.M. Ceperley, Phys. Rev. Lett. {\bf 96}, 105302 (2006).
\bibitem{note} E. Vitali, M. Rossi, F. Tramonto, D.E. Galli and L. Reatto, Phys. Rev. B {\bf 77}, 180504(R) (2008).
\bibitem{note2} M. Rossi, E. Vitali, D.E. Galli and L. Reatto, unpublished.

\end{thebibliography}
\end{document}